\documentclass[singlecolumn,epj
c3]{svjour3}
\smartqed  
\RequirePackage{graphicx}
\journalname{Eur. Phys. J. C}
\begin{document}

\title{Magnetized dark energy and the late time acceleration}


\author{Anil Kumar Yadav\thanksref{e1,addr1}
        \and
        Farook Rahaman\thanksref{e2,addr2}
        \and
        Saibal Ray\thanksref{e3,addr3}
        \and
        G.K. Goswami\thanksref{addr4}.
}

\thankstext{e1}{e-mail: abanilyadav@yahoo.co.in}
\thankstext{e2}{e-mail: rahaman@iucaa.ernet.in}
\thankstext{e3}{e-mail: saibal@iucaa.ernet.in}

\institute{Department of Physics, Anand Engineering College,
Keetham, Agra 282 007, India\label{addr1}
          \and
          Department of Mathematics, Jadavpur University, Kolkata 700 032,
West Bengal, India\label{addr2}
          \and
          Department of Physics, Government College of Engineering \&
Ceramic Technology, Kolkata 700 010, West Bengal,
India\label{addr3}
          \and
           Department of Mathematics, Kalyan P. G. College, Bhilai 490 006,
India\label{addr4} }

\date{Received: date / Accepted: date}

\maketitle

\begin{abstract}
In the present work we have searched the existence of the late
time acceleration of the Universe. The matter source that is
responsible for the late time acceleration of the Universe
consists of cosmic fluid with the equation of state parameter
$\omega = p/\rho$ and uniform magnetic field of energy density
$\rho_{B}$. The study is done here under the framework of
spatially homogeneous and anisotropic locally rotationally
symmetric (LRS) Bianchi-I cosmological model in the presence of
magnetized dark energy. To get the deterministic model of the
Universe, we assume that the shear scalar $(\sigma)$ in the model
is proportional to expansion scalar $(\theta)$. This condition
leads to $A=B^{n}$, where $A$ and $B$ are metric functions and $n$
is a positive constant giving the proportionality condition
between shear and expansion scalar. It has been found that the
isotropic distribution of magnetized dark energy leads to the
present accelerated expansion of the Universe and the derived
model is in good agreement with the recent astrophysical
observations. The physical behavior of the Universe has been
discussed in details.
\end{abstract}

\section{Introduction}\label{intro}
The cosmological observations from type Ia supernovae
\cite{Riess1998,Perlmutter1999}, cosmic microwave background (CMB)
and clusters of galaxies \cite{Pope2004} etc, all suggest that the
expansion of the present Universe is speeding up rather than
slowing down. This at once indicates that the baryon matter
component is about $4\%$ of the total energy density and about
$96\%$ of the energy density in the Universe is invisible which
opposes the self attraction of matter and causes the observed
expansion of the Universe to accelerate. This acceleration is
characterized by the negative pressure and the positive energy
density and hence violates the strong energy condition. This
violation gives a reverse gravitational effect. Due to this
effect, the Universe gets a jerk and the transition from the
earlier deceleration phase to the recent acceleration phase takes
place \cite{Caldwell2006}.

In physical cosmology and astrophysics, the simplest candidate for
the dark energy (DE) is the cosmological constant $(\Lambda)$
\cite{Carroll2001,Peebles2003}. However, it needs to be extremely
fine-tuned to satisfy the current value of the DE density, which
is a serious problem \cite{Overduin1998}. Alternatively, to
explain the decay of the density, the different forms of
dynamically changing DE with an effective equation of state (EoS),
$\omega = p/\rho < -1/3$, were proposed instead of the constant
vacuum energy density. There are several other possible forms of
DE such as quintessence $(\omega
> - 1)$ \cite{Steinhardt1999}, phantom $(\omega < - 1)$
\cite{Caldwell2002} etc. It is observed that current cosmological
data from SN Ia (Supernovae Legacy Survey, Gold sample of Hubble
Space Telescope) \cite{Riess2004,Astier2006}, CMBR (WMAP,
BOOMERANG) \cite{Eisentein2005,MacTavish2006} and large scale
structure (Sloan Digital Sky Survey) \cite{Komatsu2009} do not
support possibility of $\omega << - 1$. However, time-dependent DE
characterized by $\omega = - 1$ and crossing the phantom divide
line even now is a favorable candidate. The limit obtained from
observational results coming from CMBR anisotropy and galaxy
clustering is  $-1.44 < \omega < -0.92$ with $68\%$ confidence
level \cite{Komatsu2009,Hinshaw2009}.

Under the above circumstances, it is observed that in recent years
Bianchi universes have been gaining an increasing interest and
tremendous impetus of observational cosmology. In connection to
the WMAP data \cite{Hinshaw2009,Hinshaw2003,Jaffe2005} it is now
revealed that the standard cosmological model requires a positive
and dynamic cosmological parameter that resembles the Bianchi
morphology
\cite{Jaffe2006a,Jaffe2006b,Campanelli2006,Campanelli2007}.
According to this, the Universe should achieve the following
features: (i) a slightly anisotropic special geometry in spite of
the inflation, and (ii) a nontrivial isotropization history of
Universe due to the presence of an anisotropic energy source. The
anomalies found in the cosmic microwave background (CMB) and large
scale structure observations stimulated a growing interest in
anisotropic cosmological model of Universe. Here we confine
ourselves to model LRS Bianchi-I whose spatial sections are flat
but the expansion or contraction rate are direction dependent. For
studying the possible effects of anisotropy in the early Universe
based on the present day observations many researchers
\cite{Huang1990,Chimento1997,Lima1996,Lima1994,Pradhan2004,Saha2006a,Saha2006b}
have investigated Bianchi type-I models from different point of
view. We notice that, in connection to anisotropic equation of
state for dark energy, several works are now available in the
literature \cite{Richard2009,Campanelli2010,Stephen2011}. Some
Authors
\cite{Akarsu2010,Kumar2011,Amirhashchi2011,Yadav2011a,Yadav2011b,Yadav2011c,Yadav2011d}
have studied anisotropic DE models even with constant deceleration
parameter (DP).

We would like to further mention that unlike the FRW model this
Bianchi-I type model describes a different kind of Universe in
which the scale factor is not restricted to be the same in each
direction. In the present work, following King and Coles
\cite{King2007}, we assume a large scale homogeneous magnetic
field which is responsible for the anisotropy in the flat
Universe. It is expected that such a magnetic field will impose a
single preferred direction in space. Therefore, under the
influence of this directional magnetic field along the field lines
the anisotropy of the spacetime will be axisymmetric. However, it
is argued by King and Coles \cite{King2007} that the observed
level of isotropy in the CMB places tight constraints upon the
strength of any Hubble scale magnetic field. The upper limit on
such a field is found to be of the order of $10^{-9}$ Gauss
\cite{Barrow1997}. King and Coles \cite{King2007} also state that
these and other similar limits quote the adiabatically expanded,
present-day equivalent very weak value of the field strength which
equate to a much stronger field at very early times.

It is a obvious question then - does the above size of the
magnetic field satisfy all observational constraints at the
different scales? An extensive field survey provides that (i) the
lowest measured intergalactic fields and close to the
observational upper limits via Faraday rotation measurements
\cite{Kim1991,Perley1991,Kronberg1994}, may well be of
cosmological origin; (ii) a similar protogalactic field strength
is inferred from the detection of fields of order $10^{-6}$ Gauss
in high redshift galaxies \cite{Kronberg1992} and in damped Lyman
alpha clouds \cite{Wolfe1992}, (iii) primordial nucleosynthesis
constraints only limit the equivalent current epoch field to be
less than about $10^{-7}$ Gauss \cite{Grasso1995}, a value that is
only slightly stronger than the dynamical constraint at
nucleosynthesis \cite{Thorne1967,Doroshkevich1967,Jacobs1969}.

Regarding the impact of magnetic field it is argued that in the
early times, the magnetic field had the significant role on the
dynamics of the Universe depending on the direction of the field
lines \cite{Maden1989,King2007}. Several authors have used Bianchi
models to investigate the influence of the magnetic field on the
evolution of the Universe. It is worth noting that there has been
some work on magnetic fields in Bianchi I models in the past
\cite{Jacobs1969,Milaneschi1985,LeBlanc1997,Tsagas2000}.
Milaneschi and Fabbri \cite{Milaneschi1985} studied the anisotropy
and polarization of CMB radiation where as, Jacobs
\cite{Jacobs1969} explored the effect of a uniform primordial
magnetic field. Both the investigating groups used Bianchi-I model
of the Universe. Jacobs \cite{Jacobs1969} argued that in the early
stages of the evolution of the Universe, the magnetic field
produced large expansion anisotropies during the
radiation-dominated phase whereas it has negligible effect during
the dust-dominated phase.

In the relatively recent works, it is seen that King and Coles
\cite{King2007} have used the magnetized perfect fluid
energy-momentum tensor to discuss the effects of magnetic field on
the evolution of Universe. Sharif and Zubair \cite{Sharif2010}
have studied dynamics of Bianchi-I universe with magnetized field
of anisotropic dark energy.

In the present work, however, we present a magnetized dark energy
model with time varying DP in LRS~Bianchi-I spacetime. The
investigation is organized as follows: The metric and field
equations are presented in section 2. Section 3 deals with the
exact solutions of field equations and physical behavior of the
model. The comparison between distance modulus $(\mu)$ of derived
model and observational $\mu(z)$ is presented in section 4
respectively. Finally the results are discussed in section 5.

\section{The Metric and Field  Equations}
We consider the LRS Bianchi type I metric of the form
\begin{equation}
\label{eq1} ds^2 = -dt^2 + A(t)^2dx^2 + B(t)^2 \left(dy^2 +
dz^2\right).
\end{equation}
where, A and B are functions of $t$ only. In the limit where $A(t)
= B(t)$, the metric equation (\ref{eq1}) reduces to flat FRW
metric. Here the geometry of space-time (1) is represented by two
equivalent transverse directions $y$ and $z$ and one different
longitudinal direction $x$, along which the magnetic field is
oriented i. e. $A = A_{x}$ only.

The Einstein's field equations, in the units $8\pi G=c=1$,  read
as
\begin{equation}\label{eq2}
R^{i}_{j}-\frac{1}{2} g^{i}_{j}R =- T^{i}_{j},
\end{equation}
where $T^{i}_{j}$ is the energy momentum tensor cosmic fluid and
it is given by
\begin{equation}
\label{eq3}
T^{i}_{j}=diag[\rho-\rho_{B},-\omega\rho+\rho_{B};,-\omega\rho-\rho_{B},-\omega\rho-\rho_{B}],
\end{equation}
where $\rho$ is the energy density of cosmic fluid, $\rho_{B}$ is
the energy density of magnetic field, $\omega = p/\rho$ is the EoS
parameter of cosmic fluid and $p$ is the pressure of the cosmic
fluid. It is important to note here that the EoS parameter
$(\omega)$ is not necessarily constant \cite{Carroll2003}.

It is to note that the energy momentum for cosmic fluid given in
Eq. (3) is a system of perfect fluid and magnetic field in a
comoving coordinates i.e. $T_i^j = T_i^j(PF)  +E_i^j$. Here,
$E_i^j$ is the electromagnetic field. We choose the magnetic field
along $x$-direction. In our model, the electromagnetic field
tensor $F^{ij}$ has only one non-vanishing component, viz.,
$F_{yz} =$ constant. Therefore, for the electromagnetic field
tensor $E^i_j$, one gets the following non-trivial components,
$E_t^t = E_x^x =-E_y^y =-E_z^z = \rho_B$.

Therefore, the Einstein's field equations (\ref{eq2}) for the
line-element (\ref{eq1}) reduce to the following system of
equations
\begin{equation}
\label{eq4} 2\frac{\ddot{B}}{B} + \frac{\dot{B}^2}{B^2}  =
-\omega\rho + \rho_{B}\;,
\end{equation}
\begin{equation}
\label{eq5} \frac{\ddot{A}}{A} + \frac{\ddot{B}}{B} +
\frac{\dot{A}\dot{B}}{AB} = -\omega\rho - \rho_{B} \;,
\end{equation}
\begin{equation}
\label{eq6} \frac{\dot{B}^2}{B^2} + 2\frac{\dot{A}\dot{B}}{AB}
 = \rho + \rho_{B}.
\end{equation}
Here, and in what follows, over-dots indicates differentiation
with respect to $t$. The energy conservation equations
$T^{i}_{j;i}=0$, related to the two equations for cosmic fluid and
magnetic field \cite{King2007}, are as follows
\begin{equation}
\label{eq7} \dot{\rho}+3(1+\omega)\rho H = 0,
\end{equation}
\begin{equation}
\label{eq8} \rho_{B}=\frac{\beta}{B^{4}}.
\end{equation}
Here $\beta$ is positive constant and $H$ is the mean Hubble
parameter, which for LRS Bianchi-I space-time can be defined as
\begin{equation}
\label{eq11}
H=\frac{\dot{a}}{a}=\frac{1}{3}\left(\frac{\dot{A}}{A}+2\frac{\dot{B}}{B}\right),
\end{equation}
with $a$ being the average scale factor of LRS Bianchi type-I
model and can be expressed as
\begin{equation}
 \label{eq9}
a=(AB^{2})^{\frac{1}{3}}.
\end{equation}

It seems that any radiation field with the evolution law expressed
in Eq. (8) would work. However, one may raise a tricky question
that why does this field only scale with the scale factor $B$? It
can be verified that if one writes the conservation equation for
cosmic fluid and the magnetic field separately, then one can get
Eq. (8) which contains only scale factor $B$.

The spatial volume (V) is given by
\begin{equation}
\label{eq10} V = a^{3} = AB^{2}.
\end{equation}
The expansion scalar ($\theta$), the shear scalar ($\sigma$) and
the mean anisotropy parameter ($A_{m}$) are defined as
\begin{equation}
\label{eq12} \theta =3H = \frac{\dot{A}}{A}+2\frac{\dot{B}}{B},
\end{equation}
\begin{equation}
\label{eq13}
 \sigma^{2}=\frac{1}{2}\left(\sum_{i=1}^{3} H_{i}^{2}-\frac{1}{3}\theta^{2}\right),
\end{equation}
\begin{equation}
\label{eq14} A_{m} =
\frac{1}{3}\sum_{i=1}^{3}\left(\frac{H_{i}-H}{H}\right)^{2},
\end{equation}
where $H_{i}(i = 1, 2, 3)$ represent the directional Hubble
parameters in the direction of $x$, $y$ and $z$ respectively.

\section{Solutions to the Field Equations}
In order to solve the field equations completely, we constrain,
the system of equations with proportionality relation of shear
$(\sigma)$ and expansion scalar $(\theta)$. This condition leads
to the following relation between metric potentials
\begin{equation}
\label{eq15} A=B^{n},
\end{equation}
where $n$ is the positive constant. For anisotropic model $n \neq
1$.

The motivation behind the assumption, given in the Eq.
(\ref{eq15}), can be explained with reference to the work of
Thorne \cite{Thorne1967}. The observations of the
velocity-redshift relation for extragalactic sources suggest that
Hubble expansion of the Universe is isotropic today within
approximately $30$ percent or to put more precisely, redshift
studies place the limit $\sigma/H \leq 0.3 $ on the ratio of shear
($\sigma$) to Hubble constant ($H$) in the neighborhood of our
galaxy today \cite{Kantowski1966,Kristian1966}. In this connection
it is also to be mentioned that Collins et al. \cite{Collins1980}
have pointed out that for LRS type spatially homogeneous
spacetime, the normal congruence to the homogeneous hypersurfaces
satisfy the condition $\sigma/\theta$ as constant which leads to
the assumption $A =B^{n}$.

Eqs. (\ref{eq4}), (\ref{eq5}) and (\ref{eq15}) lead to
\begin{equation}
\label{eq16}
\frac{\ddot{B}}{B}+(n+1)\frac{\dot{B}^{2}}{B^{2}}+\frac{2\beta}{(n-1)B^{4}}=0.
\end{equation}
The general solution of the Eq. (\ref{eq16}) is given by
\begin{equation}
\label{eq17}
t=\int{\frac{dB}{n_{1}B^{-(n+1)}\sqrt{\frac{2\ell}{(k+1)}B^{\frac{k+1}{n_{1}}}+c_{1}}}},
\end{equation}
where $\ell=\frac{\beta}{n_{1}(1-n)}$, $k=n_{1}(n-2)$,
$n_{1}=\frac{1}{n+2}$ and $c_{1}$ is the constant of integration.

\begin{figure}[tbp]
\begin{center}
\includegraphics[width=0.5\textwidth]{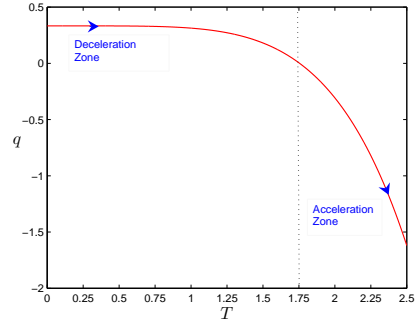} \caption{Plot of deceleration parameter $q$ versus time
$T$.} \label{fig:1.eps}
\end{center}
\end{figure}

\begin{figure}[tbp]
\begin{center}
\includegraphics[width=0.5\textwidth]{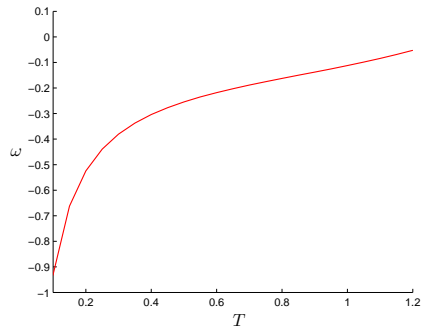}
\caption{Plot of EoS parameter $\omega$ versus time $T$.}
\label{fig:2.eps}
\end{center}
\end{figure}

Hence the spacetime (\ref{eq1}) is reduced to
\begin{equation}
\label{eq18}
ds^{2}=-\frac{dB^{2}}{n_{1}^{2}B^{-2(n+1)}\left(\frac{2\ell}{(k+1)}B^{\frac{k+1}{n_{1}}}+c_{1}\right)}+
B^{2n}dx^{2}+B^{2}(dy^{2}+dz^{2}).
\end{equation}
After using the suitable transformation of coordinates, $B=T$, the
above model (\ref{eq18}) transforms to
\begin{equation}
\label{eq19}
ds^{2}=-\frac{dT^{2}}{n_{1}^{2}T^{-2(n+1)}\left(\frac{2\ell}{(k+1)}T^{\frac{k+1}{n_{1}}}+c_{1}\right)}+
T^{2n}dx^{2}+T^{2}(dy^{2}+dz^{2}).
\end{equation}

It is to be noted that Eq. (\ref{eq17}) indicates the explicit
dependence of $B$ on $t$, i.e. $B = B(t)$. However, one can not
solve Eq. (\ref{eq17}) in general. So, in order to solve the
problem completely, we have to choose either $B$ or $n$ in such a
manner that (\ref{eq17}) be integrable. It can be easily checked
that different suitable values of $n$ generate the numerical
solution of Eq. (\ref{eq17}). But we are looking for a physically
viable model of Universe and this prompt us to consider the
transformation $B = f(t) = T$, which gives the time dependent DP.
With this transformation Eq. (\ref{eq17}) leads $t$ = constant for
$T = 0$.

Now, Eq. (\ref{eq15}) yields
\begin{equation}
\label{eq20} A=T^{n}.
\end{equation}
It is important to note here that the derived model recovers
isotropy with $n = 1$. However, when we put $n = 1$ in Eq.
(\ref{eq16}), it leads to a singularity. Therefore one can not
choose $n = 1$ to describe the feature of Universe in the present
model.

The physical parameters such as the directional Hubble's
parameters $(H_{x}, H_{y}, H_{z})$, the average Hubble parameter
$(H)$, the expansion scalar $(\theta)$, the spatial volume $(V)$
and the scale factor $(a)$ are, respectively given by
\begin{equation}
 \label{eq21}
H_{x} =
nn_{1}T^{-(n+2)}\sqrt{\frac{2\ell}{(k+1)}T^{\frac{(k+1)}{n_{1}}}+c_{1}},
\end{equation}

\begin{equation}
 \label{eq22}
H_{y}=H_{z}=
n_{1}T^{-(n+2)}\sqrt{\frac{2\ell}{(k+1)}T^{\frac{(k+1)}{n_{1}}}+c_{1}},
\end{equation}

\begin{equation}
 \label{eq23}
H=\frac{1}{3}T^{-(n+2)}\sqrt{\frac{2\ell}{(k+1)}T^{\frac{(k+1)}{n_{1}}}+c_{1}},
\end{equation}

\begin{equation}
 \label{eq24}
\theta = 3H =
T^{-(n+2)}\sqrt{\frac{2\ell}{(k+1)}T^{\frac{(k+1)}{n_{1}}}+c_{1}},
\end{equation}

\begin{equation}
 \label{eq25}
V=T^{n+2},
\end{equation}

\begin{equation}
 \label{eq26}
a=T^{\frac{n+2}{3}}.
\end{equation}

The value of DP $(q)$ is found to be
\begin{equation}
\label{eq27} q=-1+\frac{1}{3}\left[\frac{\ell
(k+3)}{k+1}T^{\frac{k+1}{n_{1}}}+c_{1}\right].
\end{equation}

The sign of $q$ indicates whether the model inflates or not. A
positive sign of $q$ corresponds to the standard decelerating
model whereas the negative sign of $q$ indicates inflation. The
recent observations of SN Ia
\cite{Riess1998,Perlmutter1999,Torny2003} and CMB anisotropies
\cite{Bennett2003} disclose that the expansion of the Universe is
accelerating at present and it was decelerating in past with a
transition redshift about $0.5$. It is therefore expected a
signature flipping in the DP for the Universe which was
decelerating in the past and is accelerating at the present time
\cite{Padmanabhan2003}. In standard cosmology the DP naturally
evolves with time just because there are many fluids, and
whichever takes over determines it at any given time. However,
recently Saha and Yadav \cite{Saha2011} presented an anisotropic
DE model with time varying DP. In the present work, Fig. 1 depicts
the variation of DP versus cosmic time as representative case with
appropriate choice of constants of integration and other physical
parameters.

The shear scalar $(\sigma)$ and the mean anisotropy parameter
$(A_{m})$ are given by
\begin{equation}
 \label{eq28}
\sigma^{2} =
\frac{1}{18T^{2(n+2)}}\left[\left(1-\frac{1}{3nn_{1}}\right)^{2}+2\left(1-\frac{1}{3n_{1}}\right)^{2}\right]
\left({\frac{2\ell}{(k+1)}T^{\frac{(k+1)}{n_{1}}}+c_{1}}\right)
\end{equation}

\begin{figure}[tbp]
\begin{center}
\includegraphics[width=0.5\textwidth]{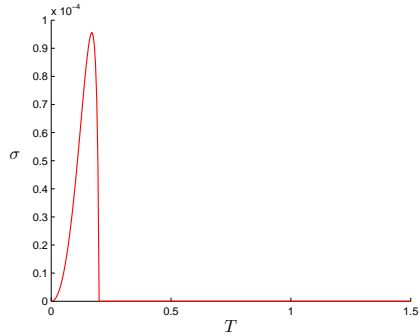}
\caption{Plot of shear scalar $\sigma$ versus time $T$.}
\label{fig:2.eps}
\end{center}
\end{figure}

\begin{equation}
 \label{eq29}
A_{m} =
\frac{1}{3}\left[\left(1-\frac{1}{3nn_{1}}\right)^{2}+2\left(1-\frac{1}{3n_{1}}\right)^{2}\right]
\end{equation}
The energy density of the cosmic fluid $(\rho)$, the EoS parameter
$(\omega)$ and the energy density of magnetic fluid $(\rho_{B})$
are found to be
\begin{equation}
 \label{eq30}
\rho=\frac{n_{1}^{2}\left(\frac{2\ell}{k+1}
T^{\frac{k+1}{n_{1}}}+c_{1}\right)}{T^{2n+2}}
\left[1+\frac{2n}{T}-\frac{\beta}{T^{4}}\right],
\end{equation}
\begin{equation}
 \label{eq31}
\omega= -\frac{\left[2n_{1}\ell
T^{\frac{k+1}{n_{1}}-1}+(3T-2)T^{2n+1}- \beta
T^{2n-2}\right]}{n_{1}^{2}\left(\frac{2\ell}{k+1}T^{\frac{k+1}{n_{1}}}+c_{1}\right)
\left(1+\frac{2n}{T}-\frac{\beta}{T^{4}}\right)},
\end{equation}
\begin{equation}
 \label{eq32}
\rho_{B} = \frac{\beta}{T^{4}}.
\end{equation}
The parameters $H$, $\theta$ and $\sigma^2$ start off with
extremely large values and continue to decrease with the expansion
of the Universe whereas the spatial volume $(V)$ grows with the
cosmic time. Fig. 2 shows the variation of EoS parameter
$(\omega)$ versus cosmic time for accelerating phase of the
Universe as a representative case with appropriate choice of
constants of integration and other physical parameters.

In the derived model the shear is limited to about $10^{-4}$ (Fig.
3) which is in fair agreement with the work of Adamek et al
\cite{Adamek2011}.

\section{Some Observational Constraint}
In this section we follow the maximum likelihood approach under
which one minimizes $\chi^{2}$ and hence measures the deviations
of the theoretical predictions from the observations.

Let us now provide the scale factor $a$ and redshift $z$ are
connected through the relation
\begin{equation}
 \label{eq33}
a=\frac{a_{0}}{(1+z)},
\end{equation}
where $a_{0}$ is the present value of scale factor.

Combining Eqs. (\ref{eq23}), (\ref{eq26}) and (\ref{eq33}), one
can easily obtain the expression for the Hubble's parameter $(H)$
in terms of redshift parameter $(z)$ as follows
\begin{equation}
 \label{eq34}
H=H_{0}(1+z)^{\frac{3(n+4)}{n+2}}.
\end{equation}
Here $H_{0}$ is the present value of the Hubble's parameter. Note
that this relation is obtained by omitting the constant of
integration.

\begin{figure}
\begin{center}
\includegraphics[width=4.0in]{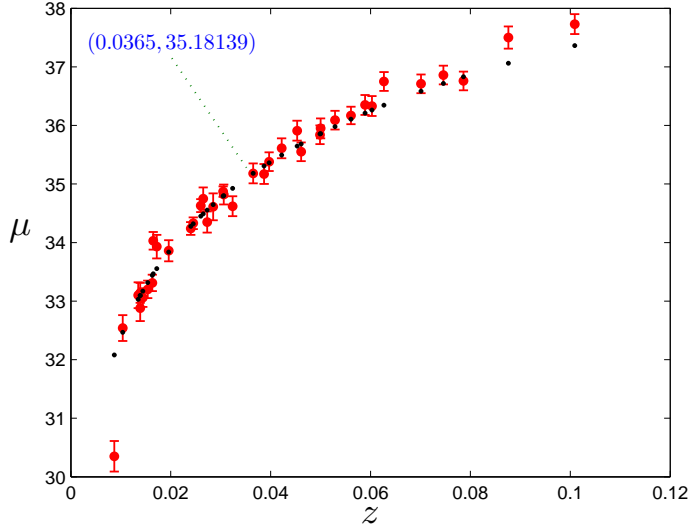}
\caption{The distance modulus $\mu$ as function of redshift (z) to
the derived model compared with SN Ia data from Amanullah et al.
\cite{Amanullah2010}. The observational $\mu(z)$ data points are
shown with error bars and the solid dots corresponds to distance
modulus of derived model.} \label{fig:4.eps}
\end{center}
\end{figure}

It is necessary for the investigation of type Ia supernovae to
explore DE and constraint the models. Since SN Ia behave as
excellent standard candles, they can be used to directly measure
the expansion rate of the Universe upto high redshift, comparing
with the present rate. The SN Ia data gives us the distance
modulus $(\mu)$ to each supernova as
\begin{equation}
\label{eq37} \mu = 5log_{10}D_{L}(z) +\mu_{0},
 \end{equation}
where $D_{L}=\frac{H_{0}d_{L}}{c}$ is the Hubble-free luminosity
distance and $\mu_{0}$ is the zero point offset, defined as
\begin{equation}
 \label{eq38}
\mu_{0}=5log_{10}\left(\frac{cH_{0}^{-1}}{Mpc}\right) +25.
\end{equation}
Inserting Eq. (\ref{eq38}) into Eq. (\ref{eq37}), we obtain
\begin{equation}
 \label{eq39}
\mu=5log_{10}\left(\frac{d_{L}}{Mpc}\right) +25.
\end{equation}
The luminosity distance $(d_{L)}$ is calculated by
\begin{equation}
 \label{eq40}
d_{L}=r_{1}(1+z)a_{0}.
\end{equation}

For the determination of $r_{1}$, we assume that a photon emitted
by a source with co-ordinate $(r,t)$ and received at a time
$t_{0}$ by an observer located at $r = 0$. Then we determine
$r_{1}$ from following relation
\begin{equation}
\label{eq41} r_{1} = \int^{t_{0}}_{t}\frac{dt}{a}.
\end{equation}
By solving the Eqs. (\ref{eq39})$-$(\ref{eq41}) and (\ref{eq26}),
one can easily obtain the expression for distance modulus $(\mu)$
in the term of red shift parameter $(z)$ as
\begin{equation}
 \label{eq42}
\mu =
5log_{10}\left[\frac{3H_{0}^{\frac{1}{(n+2)k}}(1+z)^{\frac{2n+1}{n+2}}}
{{\sqrt{\frac{2\ell}{k+1}}}^\frac{1}{(n+2)k}}\left((1+z)^{\frac{1-n}{n+2}}-1\right)\right]+25,
\end{equation}
where $H_{0}$ is in the unit of Km $s^{-1}$Mp$c^{-1}$.

\begin{table}
\caption{Comparison of the results of present model with the
observational data} \label{tab:1}
\begin{tabular}{@{}lllll@{}}
\hline \hline Redshift$(z)$& Supernovae Ia$(\mu^{data})$ & Our
model $(\mu^{th})$\\ \hline 0.0087 &$30.35^{+0.26}_{-0.26}$
&32.07943\\ \hline 0.0104 &$32.54^{+0.22}_{-0.22}$ &32.46627\\
\hline 0.0135 &$33.10^{+0.22}_{-0.22}$ &33.031435\\ \hline 0.0172
&$33.93^{+0.2}_{-0.2}$  &33.555816
\\ \hline 0.0245
&$34.33^{+0.1}_{-0.01}$ &34.320871\\ \hline 0.0285 &
$34.61^{+0.23}_{-0.23}$ & 34.64755
\\ \hline 0.0306
&$34.82^{+0.17}_{-0.17}$ &34.8010\\ \hline 0.0365
&$35.18^{+0.17}_{-0.17}$ &35.18139 \\ \hline 0.0453
&$35.91^{+0.17}_{-0.17}$ &35.64668\\ \hline 0.0499
&$35.84^{+0.16}_{-0.16}$ &35.85475\\ \hline 0.0529
&$36.09^{+0.16}_{-0.16}$& 35.98026\\ \hline 0.0589
&$36.35^{+0.17}_{-0.17}$ &36.21104 \\  \hline 0.0603
&$36.33^{+0.17}_{-0.17}$ &36.26146 \\ \hline 0.0627
&$36.75^{+0.16}_{-0.16}$    &36.34521\\ \hline 0.0701
&$36.71^{+0.16}_{-0.16}$ &36.58437 \\ \hline 0.0746
&$36.86^{+0.16}_{-0.16}$ &36.71760 \\ \hline 0.0786
&$36.76^{+0.16}_{-0.16}$ &36.82936 \\ \hline  0.0876
&$37.5^{+0.19}_{-0.19}$ &37.06104\\  \hline 0.1009
&$37.73^{+0.17}_{-0.17}$ &37.36251\\  \hline
\end{tabular}
\end{table}

In the present analysis, we use $19$ data set out of recently
released $38$ data set of SN Ia in the range $0.0015 \leq z <
0.12$, as reported by Amanullah et al. \cite{Amanullah2010} (Table
1). In this case $\chi_{SN}^{2}$ has been computed according to
the following relation
\begin{equation}
 \label{eq43}
\chi_{SN}^{2} =
A-\frac{B^{2}}{C}+log_{10}\left(\frac{C}{2\pi}\right) = 97.772,
\end{equation}
where\\
$$A=\sum_{i}^{38}\frac{\left(\mu^{data}-\mu^{th}\right)^2}{\sigma_{i}^{2}},$$
$$B=\sum_{i}^{38}\frac{\left(\mu^{data}-\mu^{th}\right)}{\sigma_{i}^{2}},$$
$$C=\sum_{i}^{38}\frac{1}{\sigma_{i}^2}.$$

The comparison between distance modulus $\mu$ of the derived model
and the observational $\mu(z)$ SN Ia data as reported by Amanullah
et al. \cite{Amanullah2010} can be seen in Fig. 4. The
observational $\mu(z)$ data points are shown with error bars and
the solid dots corresponds to distance modulus of the derived
model. It is observed that the derived model is fit well with SN
Ia observation (see Fig. 4). Note that the best fit value of
distance modulus is $\mu(z\;=\;0.0365) = 35.18139$ with
$\chi_{SN}^{2} = 97.772$ and the reduced $\chi^{2}$ value is
$\chi_{SN}^{2}$/(degree of freedom)=2.64.

\section{Results and Discussions}
In this paper, we have investigated magnetized DE model under the
assumption that $\sigma \propto \theta$ in Bianchi-I spacetime.
Under some specific choice of the parameters the present
consideration yields the time dependent DP and EoS parameters. It
is to be noted that our procedure of solving the field equations
are altogether different from what Sharif and Zubair
\cite{Sharif2010} have adopted. The derived model starts expanding
with Big Bang singularity at $T=0$ and this singularity is point
type because the directional scale factors $A$ and $B$ both
simultaneously vanish at $T=0$. The dynamics of DP parameter
yields two different phases of the Universe. Initially DP is
evolving with positive sign that yields the decelerating phase of
the Universe whereas in the later times it is evolving with
negative sign which describes the present phase of the
acceleration of the Universe. Thus the derived model has
transition of the Universe from the early deceleration phase to
the later acceleration phase which is in good agreement with the
recent observations \cite{Caldwell2006}.

The distance modulus of the derived model fit well with the
observational $\mu(z)$ values (see Fig. 4 and Table 1) which in
turn imply that the derived model is physically realistic. It is
important to note here that in the absence of the magnetic field
only the anisotropic distribution of DE leads the present
acceleration of the Universe \cite{Yadav2011c} while in the
presence of magnetic field along with the isotropic distribution
of DE describes the dynamics of the Universe from Big Bang to the
present epoch. Thus the magnetic field isotropizes the
distribution of DE which signifies the role of magnetic field.
Hence from the theoretical perspective, the present model can be a
viable model to explain the late time acceleration of the
Universe. In other words, the solution presented here can be one
of the potential candidates to describe the observed Universe.

It is worth noticing that we have presented solutions of a
magnetized Bianchi-I Universe like King and Coles \cite{King2007}.
However, their work is concerned with vacuum energy. In our case,
we have studied the Universe consists of cosmic fluid with
equation of state $w = p/\rho$. Therefore, our study is more
general than King and Coles \cite{King2007}.

In the present work we observe an interesting feature that the
magnetic presence isotropises the expansion in the model. However,
it is not clear, in particular, what property of the magnetic
field is responsible for the isotropisation. Is it the additional
energy density of the field, is it the magnetic pressure, or maybe
the tension? It can be observed that the solution of Eq. (16) does
not exist for $n = 1$. Therefore, in the derived model, one can
not choose $n = 1$. To obtain an explicit solution of Eq. (16),
one can choose $\beta = 0$ which leads to $\rho_B = 0$ and the
isotropic distribution of cosmic fluid. This can also be observed
from the LHS's of Eqs. (4) and (5), which are same for $n = 1$. As
a result, one should get $\rho_B = 0$. This means, there is no
contribution of magnetic field and the model becomes an isotropic
FRW model. The well known criteria for isotropization are $A
=\frac{1}{3}\Sigma \frac{H_i^2}{H^2} - 1 \rightarrow 0$ and
$\Sigma^2 =\frac{1}{2}AH^2 - 1 \rightarrow 0$, where $A$,
$\Sigma^2$ are average anisotropy and shear respectively, and
$H_i^2$ is the directional Hubble parameter. Obviously, these
criteria are valid for large physical time. In the present model
(Eqs. (21) - (24)), one can easily find out that for large time,
$A \rightarrow \frac{2(n-1)^2}{n+2)^2}$ and $\Sigma^2 \rightarrow
0$. This immediately implies that for $n \rightarrow 1$ our model
turns out to be an isotropic FRW model. Therefore, $n =1$ is the
condition of isotropy in the absence of magnetic field and the
presence of magnetic field has constraint on $n = 1$. This seems
to contribute the magnetic field which resembles with the initial
anisotropy of the Universe. With the passage of time magnetic
field decreases and becomes negligible at late time to approach
towards isotropy.

Our solution, in the present investigation, shows that the
constrained equation of state of $\rho$ evolves in time, but this
is not all that is needed when comparing with observations. This
explicitly means that it is not exactly known about the generic
fluid $\rho$, and how does it describe and include the known
cosmological history where an early radiation domination gave way
to matter domination. In particular, fitting the $H(z)$ evolution
law derived from SN observations is only one of the many pieces of
information which need to be used. A thorough discussion of all
basic cosmological constraints is beyond the scope of this
analysis in the present paper and is awaiting for a future
project.

\section*{Acknowledgements}
AKY would like to thank The Institute of Mathematical Science
(IMSc), Chennai, India for providing facility and support where a
part of this work was carried out. We all are thankful to two
anonymous referees for their useful comments which have enabled us
to improve the manuscript substantially.

\end{document}